\documentclass[11pt]{article}
\usepackage[utf8]{inputenc}
\usepackage{amsmath}
\usepackage{wrapfig}
\usepackage{graphicx}
\usepackage{float}
\usepackage{caption}
\usepackage{authblk}
\usepackage[super,sort]{natbib}
\usepackage{amssymb}
\usepackage{eso-pic} 
\usepackage{helvet}
\usepackage{xspace}

\definecolor{Blue}{rgb}{0,0.08,0.65}

\newcommand{\hi}{{\sc H\,i}\xspace}
\newcommand{\msun}{\mbox{$\rm{M}_\odot$}\xspace}
\newcommand{\mstar}{\mbox{$M_\star$}\xspace}

\usepackage[colorlinks=False,linkcolor = blue,citecolor=blue]{hyperref}%
\usepackage[a4paper,margin=2.2cm]{geometry}

\usepackage{soul} 

\makeatletter
\renewcommand\maketitle
  {\noindent
   {\huge\bfseries\@title}%
   \medskip\par\noindent
   {\large\bfseries\@author}%
   \hfill
   {\large\@date}%
   \bigskip\par\noindent
  }
\makeatother

\title{The role of the cosmic web in the scatter of the galaxy stellar mass - gas metallicity relation}
\author[1,2,*]{Callum T. Donnan}
\author[1]{Rita Tojeiro}
\author[3]{Katarina Kraljic}

\affil[1]{\textit{School of Physics and Astronomy, University of St Andrews, North Haugh, St Andrews, KY16 9SS, UK}}
\affil[2]{\textit{Institute for Astronomy, University of Edinburgh, Royal Observatory, Edinburgh, EH9 3HJ, UK}}
\affil[3]{\textit{Aix Marseille Univ, CNRS, CNES, LAM, Marseille, France}}

\affil[*]{callum.donnan@ed.ac.uk}

\begin{document}
\maketitle


\textbf{Understanding the relationship between the cosmic web and the gas content of galaxies is a key step towards understanding galaxy evolution. However, the impact of the cosmic web on the growth of galaxies and dark matter halos is not yet properly understood. We report a detection of the effect of the cosmic web on the galaxy stellar mass - gas phase metallicity relation of low redshift star-forming galaxies, using SDSS data. The proximity of a galaxy to a node, independently of stellar mass and overdensity, influences its gas-phase metallicity, with galaxies closer to nodes displaying higher chemical enrichment than those further away. We find a similar but significantly weaker effect with respect to filaments. We supplement our observational analysis with a study of the cosmological hydrodynamical simulation IllustrisTNG (TNG300), finding qualitative agreement with our results. Using IllustrisTNG, our results can be explained by both halo assembly bias and gas supply combining in nodes in a way that significantly modulates the metallicity of the gas, contributing to the scatter of this fundamental relation in galaxy evolution.}

\section*{}
Within the standard cosmological model, galaxy formation and evolution is understood in broad terms - driven by gravity, structure forms hierarchically, yielding the potential wells within which cold baryons gather and form stars. Modern models of the galaxy-halo connection attempt to link the evolution of halos within the standard paradigm with the observed evolution of galaxies\cite{WechslerTinker2018}. We know from observations that stellar mass is key in understanding the properties of different galaxies\cite{BlantonMoustakas2009}, and the statistical link between stellar mass and halo mass is now well established. However, even at low-redshift - where observations are abundant - the details of what drives individual galaxy pathways are not resolved. Having successfully established meaningful mean empirical relationships in extragalactic observations, efforts can now focus on explaining why galaxies deviate from mean trends.

Galaxies exist within the the large-scale structure of the Universe consisting of dense nodes connected by filaments which are separated by walls and underdense voids \cite{cosmicweb} - the so-called cosmic web. The effect of the cosmic web on the growth of halos, at fixed halo mass (i.e., halo assembly bias), is now well established in simulations \cite{sim_bias1,sim_bias2,sim_bias3,sim_bias4}, and has been detected in observations \cite{HaloAssembly}: at fixed halo mass, halos closer to nodes or filaments assembled their mass earlier than those further away. However, the question remains of if, how and why the cosmic web impacts on the baryonic properties of galaxies. Simple models of the galaxy-halo connection link the star-formation history of a galaxy to the mass assembly history of the halo – i.e., gas accretion follows dark matter accretion (with some efficiency), and star-formation rate (SFR) follows gas accretion. So one may argue, at the simplest level, that halo assembly bias regulates the stellar content of a galaxy indirectly, by modulating the halo assembly history, and there’s observational evidence in that direction: at fixed stellar mass, galaxies near filaments show redder colours, have reduced SFRs, and higher stellar metallicity \cite{GalEvCW3,GalEvCW,SDSSresults}.

However, the gas ecosystem of galaxies is complex and understanding how the cosmic web affects gas in and around galaxies is crucial to understand its overall impact on the galaxy-halo connection. Some observational studies have focused on correlating \hi gas content in galaxies with some form of cosmic web estimate, to varying success and results \cite{HI_replenish,HI}. Here, we focus for the first time on the gas-phase metallicity, which describes the relative mass of elements heavier than helium in the interstellar medium (ISM) of a galaxy. There has been some insight into the environmental effects on gas-phase metallicity before \cite{Darvish+15, Genel16, Gupta+18}, but never within the full context of the cosmic web in a wide area survey. The shape of the metallicity vs stellar mass relation (MZR) is a key observable in models of galaxy evolution, and offers important insight on baryonic evolutionary mechanisms in galaxies \cite{ZReview}. The metal content in the ISM is enriched by stellar evolution, and depleted by winds that carry metals away from a galaxy and by dillution through accretion of lower-metallicity gas from the circumgalactic medium (CGM) and the intergalactic medium (IGM), which in turn flows from the cosmic web. On the mean, gas-phase metallicity climbs steeply with stellar mass, with a scatter considerably larger than measurement uncertainties\cite{SDSSMetallicities}. Understanding what drives this scatter - i.e., why do galaxies at fixed stellar mass have different gas phase metallicities? - is likely to drive our understanding of what physical processes define the evolutionary paths of galaxies. In simulations, the scatter in the MZR has been linked to outflows and inflows  \cite{scatterMZR_2,scatterMZR}, and in observations it has been shown to correlate with SFR\cite{MannucciEtAl2010}, but the question remains on how much the factors that drive these are purely internal and stochastic, or external and modulated by the cosmic web. 
 
In this article, we study the effect of the cosmic web on the gas-phase metallicity of low-redshift galaxies in observations and in simulations. Our goal is to establish whether the cosmic web contributes to the observed scatter in the MZR and - by using simulations - to explore possible underlying physical mechanisms to that contribution. Particularly, we investigate how the cosmic web impacts on the metal and gas content of galaxies via a) its modulation of halo assembly history and b) its influence on net gas supply to galaxies. We study these two potential mechanisms in one simulation alone - IllustrisTNG - acknowledging that different cosmological hydrodynamical models may show different results.

Gas-phase metallicity probes ionised rather than neutral gas. Ionised gas is less directly linked to gas supply and gas accretion, but it bares the imprint of past stellar evolution, galactic outflows, and inflows. Studying it allows one to ask what the effect of the cosmic web is on these mechanisms. Gas-phase metallicity also has the advantage that it is more readily available, and will be measurable up to $z\sim1.4$ in surveys sufficiently dense and wide to reliably extract the cosmic web such the forthcoming Dark Energy Spectroscopic Instrument survey (DESI) \cite{DESI}. DESI and other large-scale spectroscopic experiments will provide a step-change in our understanding of the role of the cosmic web in the evolution of galaxies at low-redshift, further refining the results presented here.

\section*{The MZR in observations}

\begin{figure*}
    \centering
    \includegraphics[width=\textwidth]{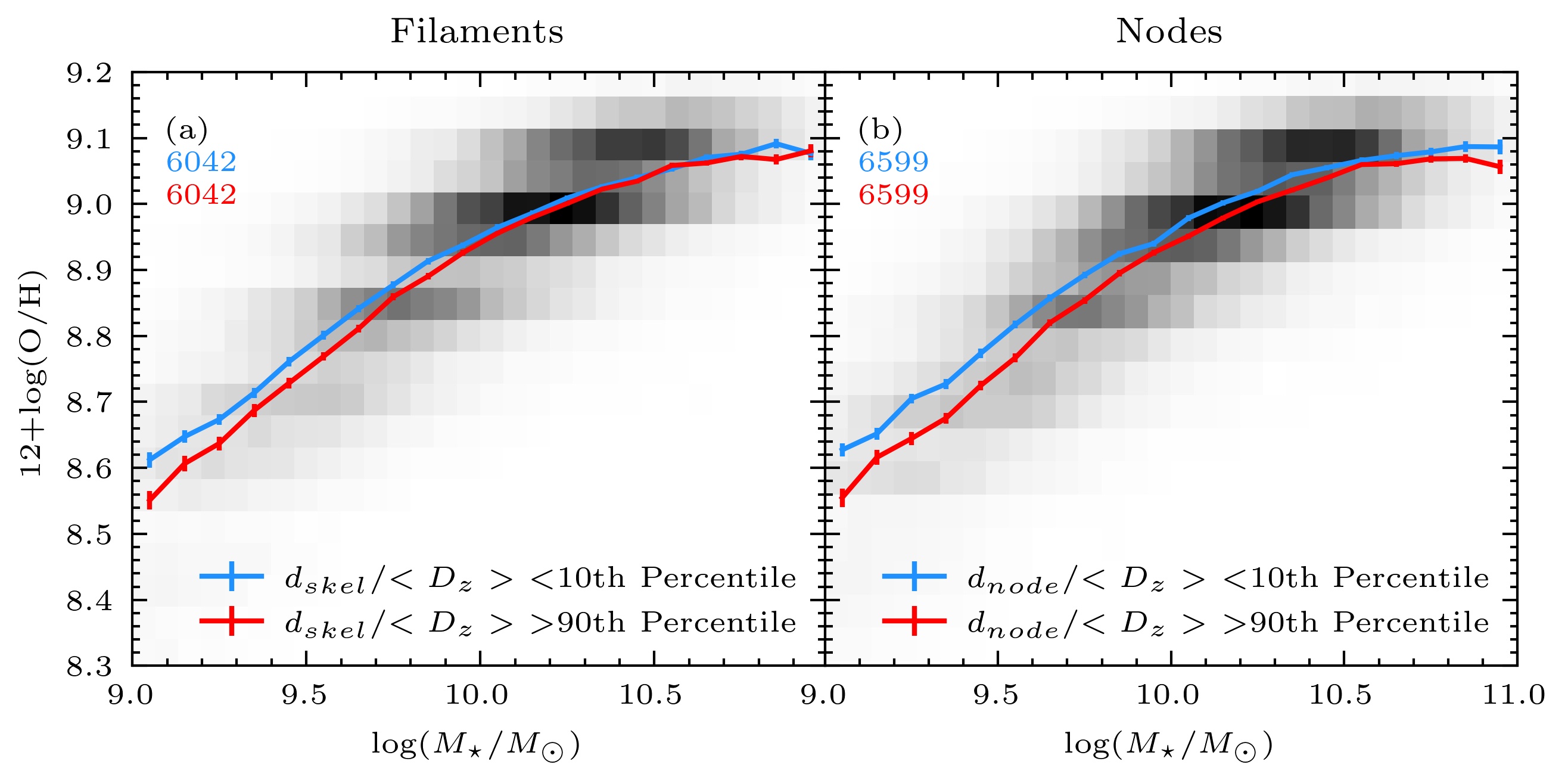}
    \caption{\textbf{Mass-metallicity relations in SDSS split by distance to filament/node.} \textit{(a)}: Mass-metallicity relation for largest 10\% of $d_{\rm skel}/\langle D_z \rangle$ and smallest 10\%, for galaxies in SDSS DR7. \textit{(b)}: Mass-metallicity relation for largest 10\% of $d_{\rm node}/\langle D_z \rangle$ and smallest 10\%, for galaxies in SDSS DR7. \textit{Both}: Solid line indicates mean gas-phase metallicity calculated for stellar mass bins of width 0.1 dex. Error bars indicate standard error on the mean. Inserted coloured text represents the number of galaxies in each MZR. Grey histogram shows 2D histogram of individual galaxies. Difference in gas-phase metallicity is present between the 10\% closest and 10\% most distant galaxies from filament and nodes, such that galaxies closer to filaments and nodes have higher gas-phase metallicities compared to their more distant counterparts.}
    \label{fig:SDSS}
\end{figure*}

Our observational sample consists of emission-line galaxies from the Sloan Digital Sky Survey\cite{SDSS_overview} (SDSS) DR7 Main Galaxy Sample\cite{sdssdr7, Strauss2002}, with a median redshift of 0.071 and median stellar mass of $10^{10}\msun$. Sample selection is detailed in the Methods section, which also provides details on the catalogue of galaxy metallicities, and estimates of distance to cosmic web features. In this study, the term cosmic web will be used to refer to regions that are topologically different, and classified as filaments or nodes (further details in Methods). There is a sampling effect as a function of redshift and therefore every distance to filaments or nodes is normalised by the redshift dependent mean inter-galaxy separation $\langle D_z \rangle$.

The MZR for this sample is shown in Fig.~\ref{fig:SDSS}, as the background histogram. As a first illustration of the effect of the cosmic web on the MZR, we begin by plotting in coloured lines the mean MZR in the 10\% of galaxies closer and further from filaments (panel a) and nodes (panel b). At fixed stellar mass, galaxies closer to filaments have higher gas-phase metallicities, compared to those more distant, particularly at low stellar mass. Galaxies closer to nodes show a higher average gas-phase metallicity,  with an average difference of approximately 0.02 dex, at all stellar masses. 

\begin{figure*}
    \centering
    \includegraphics[width=\textwidth]{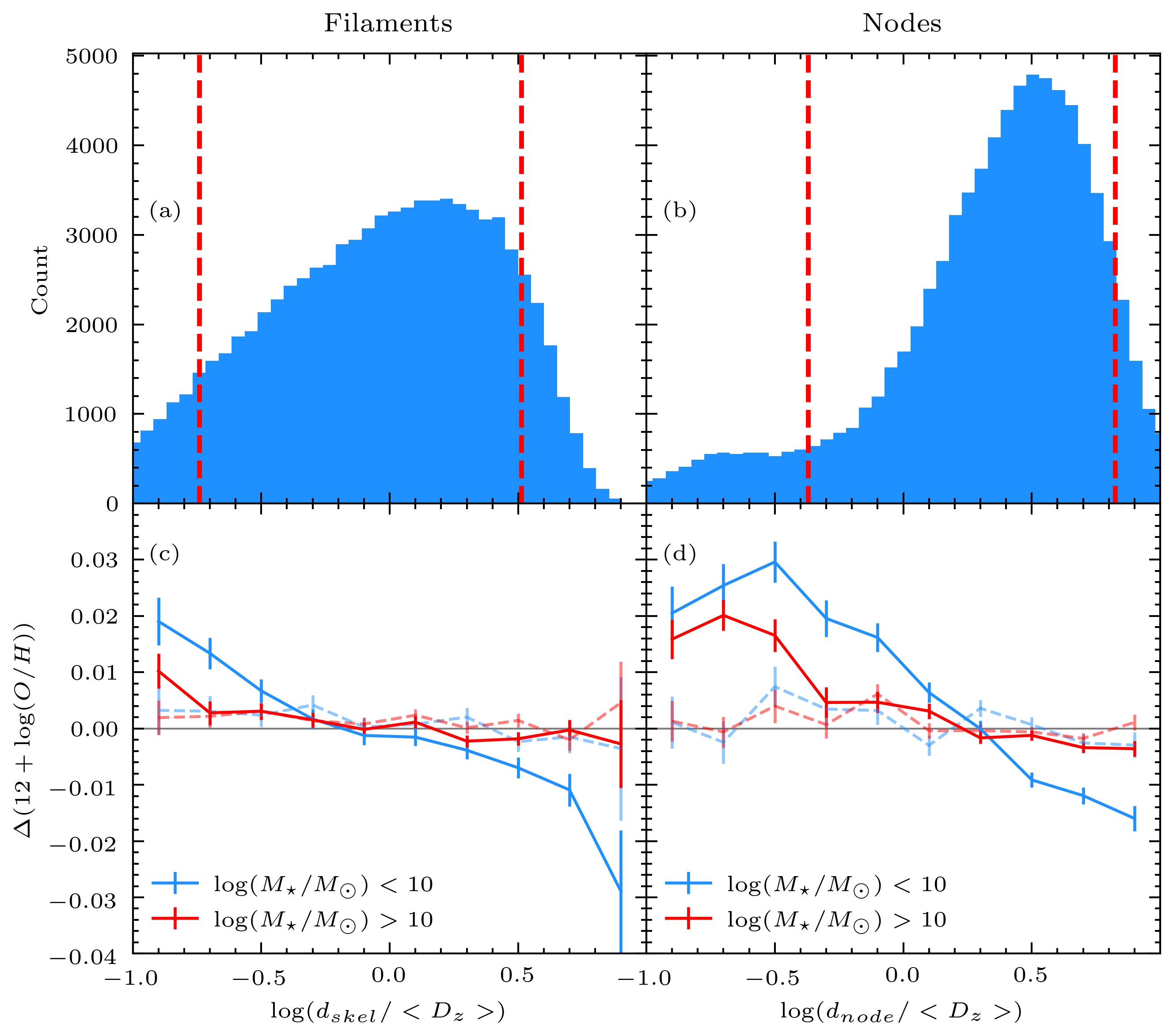}
    \caption{\textbf{Gas-phase metallicity residuals as function of distance to filament/node in SDSS.} \textit{Top}: 1-D histograms of normalised distance to filament (\textit{panel a}) and node (\textit{panel b}) for the selected SDSS galaxy sample. Dashed red lines indicate the 10th and 90th percentiles corresponding to the selection made in Fig.~ \ref{fig:SDSS}. \textit{Bottom, solid}: Binned residuals of mass-metallicity relation as a function of normalised distance to filament (\textit{panel c}) and node (\textit{panel d}). \textit{Bottom, dashed}: Binned residuals of gas-phase metallicity shuffled by overdensity as a function of normalised distance to filament (\textit{panel c}) and node (\textit{panel d}). Negative correlation bewteen gas-phase metallicity residuals and distances to node and filament (at low stellar mass) is removed when shuffling is performed.}
    \label{fig:SDSS_final}
\end{figure*}

To further quantify the effect of the cosmic web on MZR, we first compute the gas-phase metallicity residuals, with respect to the mean value of the full population, evaluated as a function of stellar mass: i.e., for each galaxy we quantify its deviation from the mean MZR at the galaxy's stellar mass. Next, we ask whether these residuals are correlated with distance to filaments or nodes. 
The top panels (a and b) of Fig.~\ref{fig:SDSS_final} show the overall distribution of distance to filaments and distance to nodes. The 10th and 90th percentiles, used in Fig.~\ref{fig:SDSS}, are indicated by the red dashed lines. The bottom panels (c and d), show how the metallicity residuals vary as a function of distance to node and filament (solid lines), in two slices of stellar mass, separated at  $\log(\mstar/\msun)=$ 10 (the median stellar mass of the sample). The correlation between gas-phase metallicity residuals and distance to nodes is negative and statistically significant in both mass bins, with the hypothesis that the two variables are uncorrelated rejected at over 5$\sigma$ (significance of the correlation between metallicity residual and distance is assessed via a Spearman rank-order coefficient; further details in Methods): {\it galaxies closer to nodes are more likely to have gas-phase metallicities above the average, when compared to those further from nodes.} We find a shallower dependence of the residuals on distance to filament, which is significant only at lower stellar masses. These observations are in agreement with the visual representation shown in Fig.~\ref{fig:SDSS}.

\subsection*{The role of overdensity}

Previous studies have shown that galaxies residing in overdensities have higher gas-phase metallicities \cite{localMetal,localMetal2,localMetal3}, and have suggested that in overdense regions the ISM was pre-enriched at higher redshift. A relationship between gas-phase metallicity and overdensity is also seen in our data (see Supplementary Fig.~3). As measurements of the cosmic web are typically correlated with overdensity\cite{Libeskind2018}, it becomes critical to assess how much our results are driven by distance to the cosmic web, {\em beyond} the known effect of overdensity. We make this assessment by shuffling our data, such that overdensity information is kept, but cosmic web distances are randomised, before recomputing the residuals (see Methods for full details). In this test, if the observed relationship between metallicity and shuffled distances remains, we can infer that it is overdensity, and not distance to cosmic web, that drives the signal in the unshuffled data. The results are shown as the faded, dashed lines in panels (c) and (d) of Fig.~\ref{fig:SDSS_final}. The gradients are visibly reduced, particularly for nodes. We therefore interpret the relationship between metallicity residual and distance to cosmic web in our unshuffled data to be primarily driven by the cosmic web, and not by overdensity. Having established  this, we  turn to IllustrisTNG to help us interpret our results.

\begin{figure*}
    \centering
    \includegraphics[width=\textwidth]{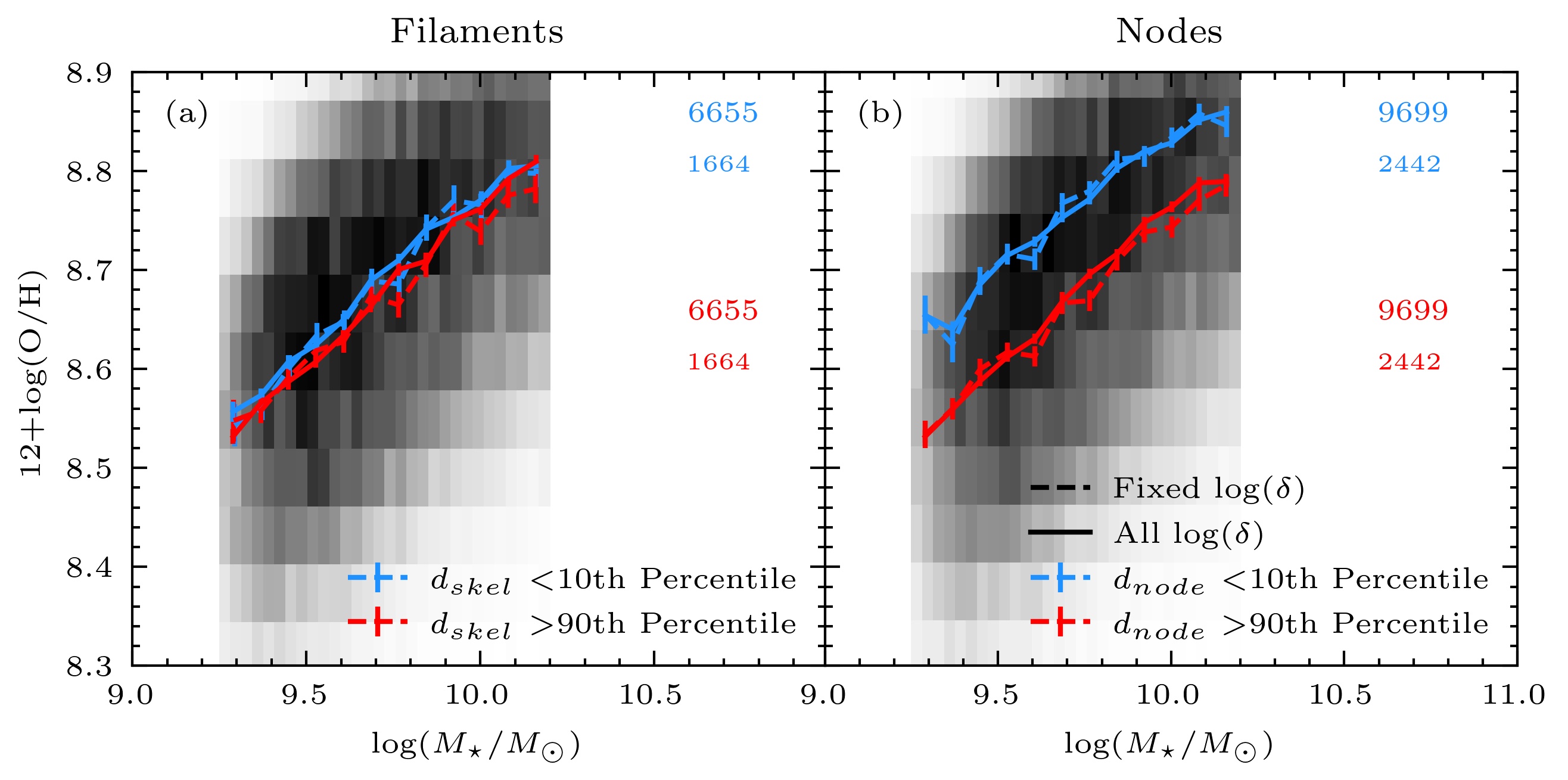}
    \caption{\textbf{Mass-metallicity relations in IllustrisTNG split by distance to filament/node.} \textit{(a)}: Mass-metallicity relations of star-forming galaxies in TNG300 at $z$ = 0.1 separated into bins by the 10\% closest and 10\% most distant to filaments. Galaxies that are at $d_{\rm node} <$ 3.5 Mpc are not included. \textit{(b)}: Mass-metallicity relations of star-forming galaxies in TNG300 separated by the 10\% closest and 10\% most distant to nodes. \textit{Both}: Solid lines indicate MZR computed for galaxies in the full overdensity range. Dashed lines indicate the MZR computed for galaxies fixed to the median overdensity ($\pm 0.32 \sigma$) in the sample. Inserted coloured text represents the number of galaxies in each MZR with the smaller number representing the number for fixed overdensity. Grey histogram shows 2D histogram of individual galaxies. Error bars indicate standard error on the mean. Difference in gas-phase metallicity is present between the 10\% closest and 10\% most distant galaxies from filament and nodes in TNG300, independently of overdensity.}
    \label{fig:TNG}
\end{figure*}

\section*{The MZR in simulations}

Modern cosmological hydro-dynamical simulations can resolve many baryonic processes on galactic scales, within a cosmological volume, and provide good matches to observations\cite{TNGColour}. Here, we use TNG300 of the IllustrisTNG project\cite{TNG, TNG300} to: (a) attempt to reproduce the trends seen in our observational work, and (b) to investigate the causes. The details of our sample selection and post-processing are given in the Methods section. In summary, we select galaxies in TNG300 by their colour, gas content and stellar mass, at $z=0.1$. We do not make mock observations in this study - our goal is not to reproduce the observations exactly, but rather to check whether simulations show the same qualitative trends and, if so, to use simulations to understand what processes may be at play.

The mass-metallicity relation for the TNG sample, split by the 10th and 90th percentiles of distance to the cosmic web is shown in the solid lines in Fig.~\ref{fig:TNG} (distance to filaments and nodes in panels a and b, respectively). There is a small difference in gas-phase metallicity with distance to filament, with a higher gas-phase metallicity found for the 10 \% closest galaxies. For nodes, a higher gas-phase metallicity for galaxies in their vicinity is clearly identified at all stellar masses in the range $9<\log(\mstar/\msun)<10.2$ with an average difference of approximately 0.1 dex between the 10\% closest and 10\% most distant galaxies. 

\begin{figure*}
    \centering
    \includegraphics[width=\textwidth]{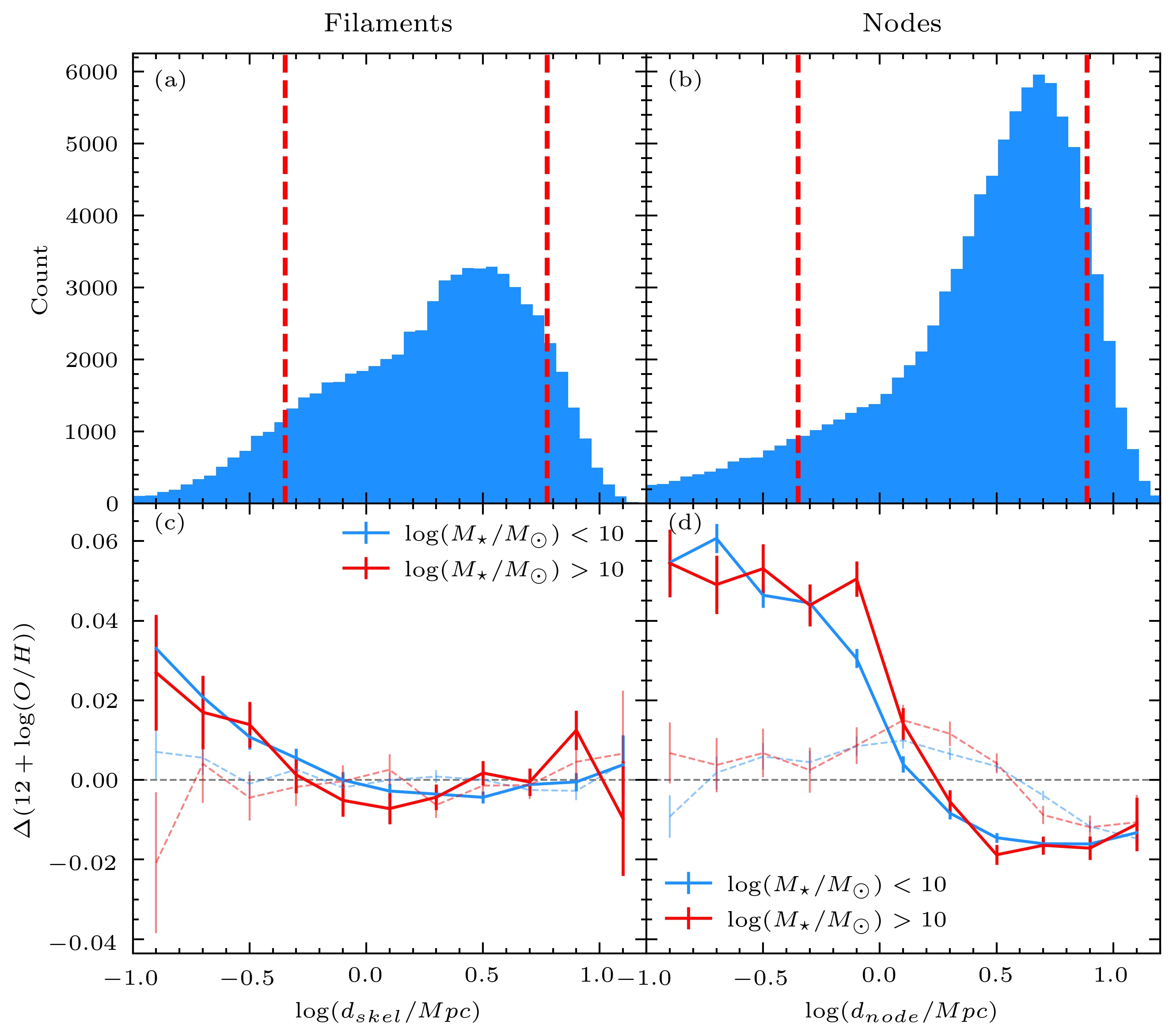}
    \caption{\textbf{Gas-phase metallicity residuals as function of distance to filament/node in IllustrisTNG.} \textit{Top}: 1-D histograms of distance to filament (\textit{panel a}) and node (\textit{panel b}) for the selected TNG300 galaxy sample. Dashed red lines indicate the 10th and 90th percentiles corresponding to the selection made in Fig.~\ref{fig:TNG}. \textit{Bottom, solid}: Binned residuals of mass-metallicity relation as a function of distance to filament (\textit{panel c}) and distance to node (\textit{panel d}). \textit{Bottom, dashed}: Binned residuals of gas-phase metallicity shuffled by overdensity as a function distance to filament (\textit{panel c}) and node (\textit{panel d}). Negative correlation between gas-phase metallicity residuals and distances to filament (up to $d_{\rm skel} \sim 10^{0.2}$ Mpc) and node is removed when shuffling is performed, pointing towards the role of cosmic web, beyond overdensity.}
    \label{fig:TNG_residual}
\end{figure*}

We investigate the correlation between gas-phase metallicity residuals and distances to the cosmic web in the same way as we did in our observational sample, shown for TNG300 in Fig.~\ref{fig:TNG_residual}. There is a strong negative correlation for nodes, with galaxies displaying a higher gas-phase metallicity than average closer to nodes. This is present at both stellar masses, with similar strength, although the sample size is significantly smaller at $\log(\mstar/\msun)>10$ due to the smaller maximum mass in the galaxy selection in Fig. \ref{fig:TNG_residual}. Once again the hypothesis that the residuals and distance to node are uncorrelated is rejected at greater than 5$\sigma$ significance using a Spearman rank-order coefficient. When looking at distances from filaments, we observe a slight trend of residual, which increases at very small distances from filaments.



To investigate the role of overdensity in the simulation, we perform the same shuffling technique as we do with the data (shown in the faded, dashed lines of panels (c) and (d) of Fig.~\ref{fig:TNG_residual}). As in the data, the correlations are not present in the shuffled data, indicating that overdensity does not dominate the signal we see in the solid lines of the same panels. In the case of TNG, we also take a more straightforward approach, afforded by the larger number of galaxies and cleaner overdensity values (enabled by the higher number density and three-dimensional positions), and we simply recompute the MZR as a function of distance to the cosmic web but using only galaxies in a small bin of overdensity, centred at the sample median $\pm 0.32 \sigma$ - this is shown by the solid lines of Fig.~\ref{fig:TNG}. The difference between the MZR computed using all overdensities and the MZR computed using a small bin of overdensity is small, and much smaller than the difference between the MZR computed at different distances from the cosmic web. We again interpret these results as evidence that the effect we observe with respect to the cosmic web is not dominated by overdensity. This conclusion holds when using overdensity computed with Gaussian kernels of 3 Mpc, 6 Mpc or 9 Mpc. 


The simulation therefore shows a qualitative agreement with observations, in that the effect is larger with nodes than filaments, and in both cases the gas-phase metallicity of galaxies increases with decreasing distance. Quantitatively, we note that the role of filaments is similar in simulations, whilst the role of nodes is amplified. This discrepancy in the role of nodes can be partly explained by the fact that the number density is higher in our TNG sample, resulting in typically smaller values of $d_{node}$. As the strength of the signal increases with decreasing $d_{node}$, we then expect a stronger signal in our TNG sample. The horizontal scales of Figures~\ref{fig:SDSS_final} and \ref{fig:TNG_residual} are not directly comparable. However, taking the mean value of $<D_z>=8$Mpc, we see that in SDSS, we have very few galaxies with $d_{node} \lessapprox 0.8$ Mpc, which is where we cut off the plot. The amplitude of the residuals at that distance is consistent between the two figures. The agreement between observations and simulations suggests that there is insight to be gained from further investigating the nature of these trends in the simulation, and we turn to that next.


\begin{figure*}
    \centering
    \includegraphics[width=\textwidth]{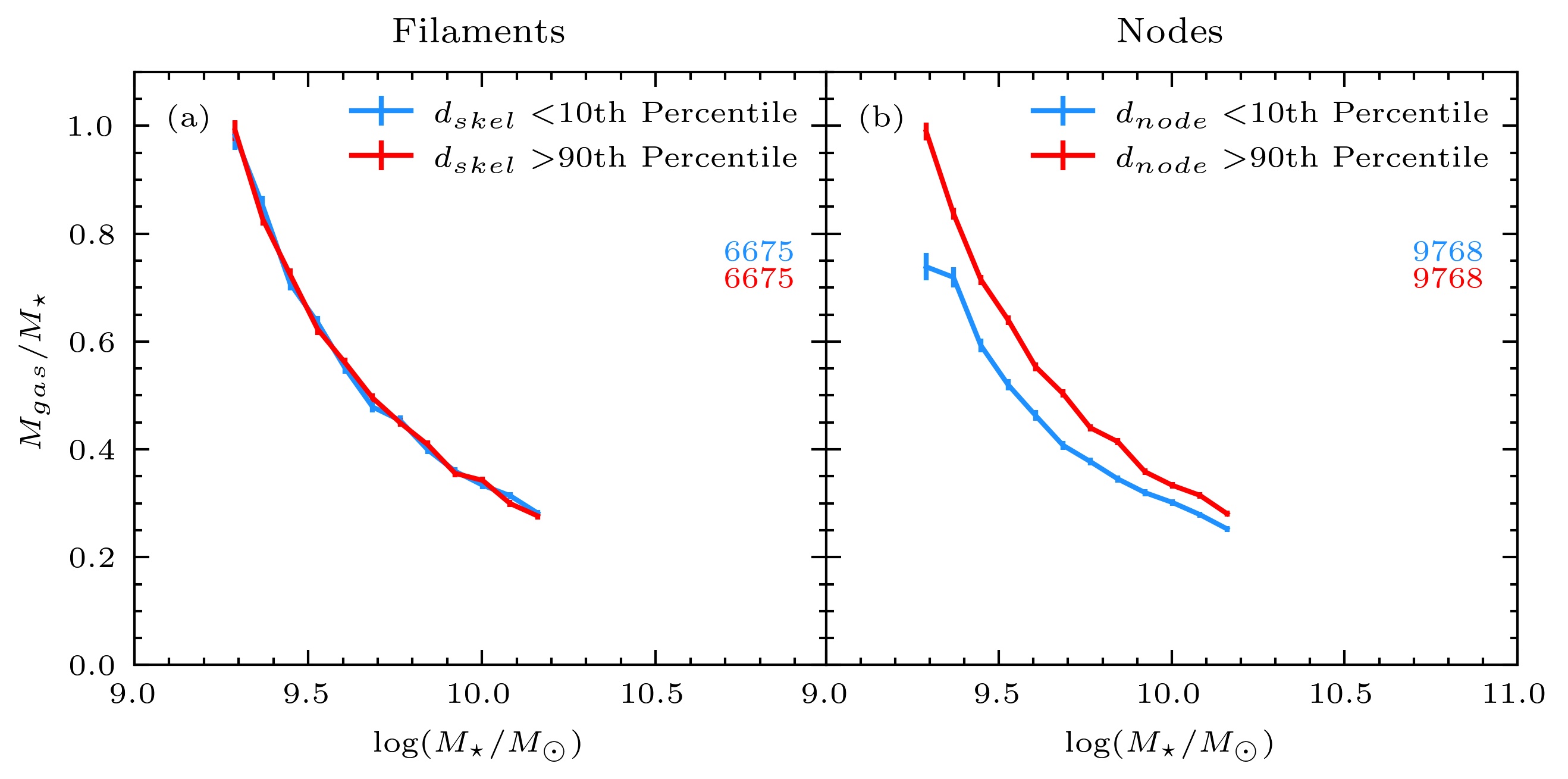}
    \caption{\textbf{Gas fraction as function of distance to filament/node in IllustrisTNG.} Gas fraction of galaxies (within stellar mass half radius) separated into the 10\% closest and 10\% most distant from filaments (\textit{a}) and nodes (\textit{b}). Error bars indicate standard error on the mean. Inserted text in colour represents the number of galaxies in each sample. Each line matches the samples shown in the respective panels of Fig.~\ref{fig:TNG}. No difference in gas fraction is found for galaxies as function of their distance to filaments. 
    Galaxies closer to nodes have lower gas fraction compared to galaxies of same mass further away from nodes.}
    \label{fig:TNG_gas_frac}
\end{figure*}

\begin{figure*}
    \centering
    \includegraphics[width=\textwidth]{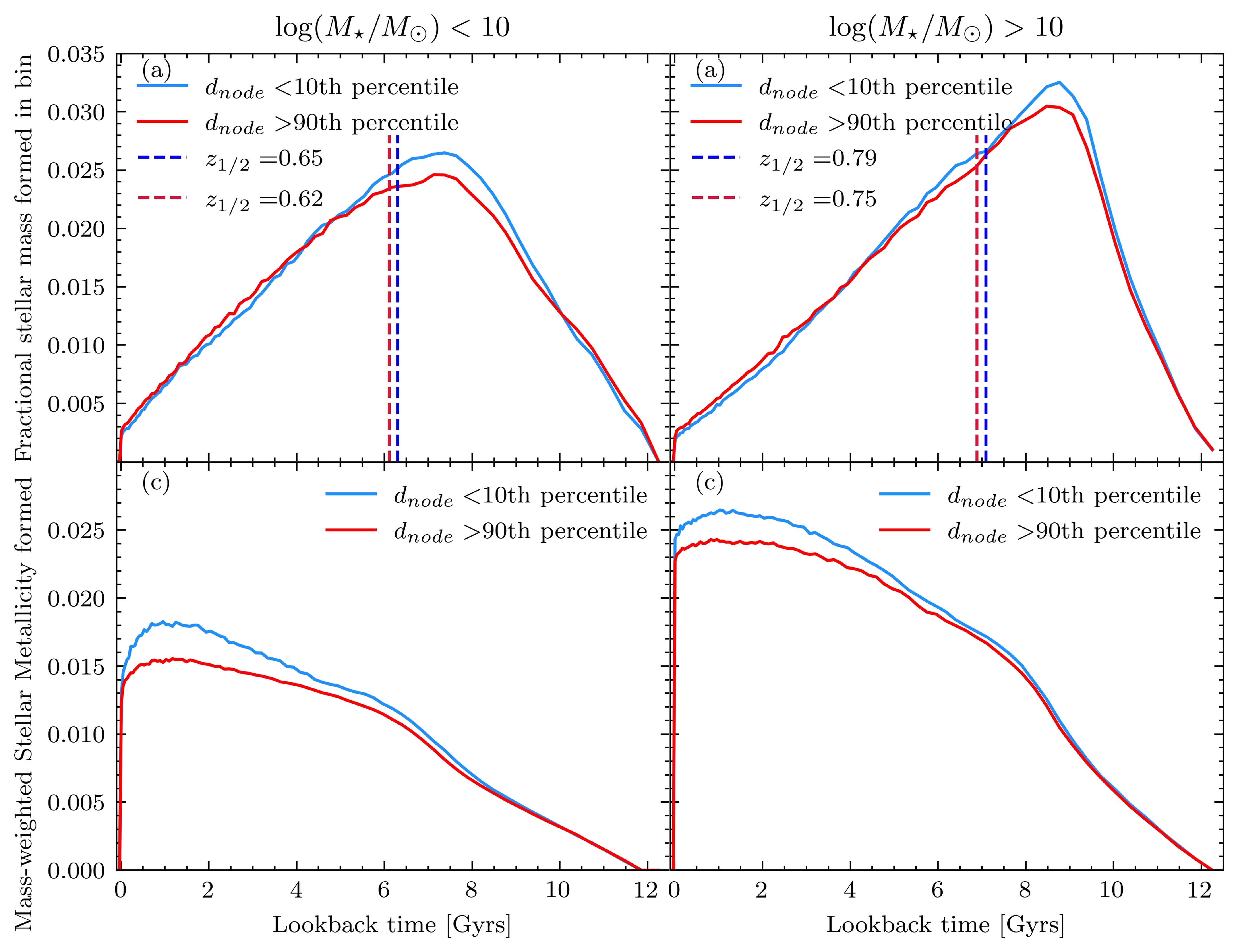}
    \caption{\textbf{Star formation and stellar metallicity histories in IllustrisTNG.} \textit{Top:} Mean fractional stellar mass formed as a function of lookback time in 400 bins for two galaxy samples at $z=0.1$, defined as the 10\% closest and 10\% most distant from nodes, at a stellar mass of $\log\left(\mstar/\msun\right)<10$ (\textit{a}) and $\log\left(\mstar/\msun\right)>10$ (\textit{b}). Dashed lines indicate the redshift at which half of the stellar mass in each sample had been assembled. \textit{Bottom:} Mean mass-weighted stellar metallicity formed in each bin as a function of lookback time in 400 bins for two galaxy samples at $z=0.1$, defined as the 10\% closest and 10\% most distant from nodes, at a stellar mass of $\log\left(\mstar/\msun\right)<10$ (\textit{c}) and $\log\left(\mstar/\msun\right)>10$ (\textit{d}). At all stellar masses considered, galaxies closer to nodes assembled their stellar mass earlier than those more distant. Galaxies closer to nodes have higher stellar metallicities particularly in their recent past.}
    \label{fig:TNG_SFH}
\end{figure*}

\section*{How does the cosmic web impact on the gas-phase metallicity of galaxies? Insights from IllustrisTNG.}
To understand how the cosmic web modulates the gas-phase metallicity, we explore different mechanisms that regulate gas-phase metallicity in galaxies. The metallicity of the gas in a galaxy is increased by star formation where stars produce metals through nuclear fusion and expel metals into the ISM through feedback such as stellar winds and supernovae. The metallicity of the gas has been theorised to be lowered by accretion of metal-poor gas from the CGM and IGM. This dilutes the enriched gas in the galaxy lowering the overall gas-phase metallicity of the ISM \cite{ZReview}. 

\subsection*{Gas mass fractions and the role of inflows}

We begin by looking at the gas fraction, $M_{\rm gas}/\mstar$, as a function of stellar mass, for galaxies in the 10th and 90th percentile of distance to nodes and filaments. The gas mass and stellar mass were computed within the stellar mass half-radius. We are, thereby, considering gas that has reached the ISM, rather than probing of gas in the circumgalactic medium \cite{CGM}. Limiting the gas to the galaxies' half-light radius is also more comparable to our observational results, due to the SDSS fiber aperture. Fig.~\ref{fig:TNG_gas_frac} displays the gas fraction as a function of stellar mass for galaxies in opposing percentiles of distance to filaments and nodes. Panel (a) shows that the difference between the gas fractions of galaxies close to filaments and those that are distant from filaments is very small; this is true at all stellar masses considered. In the scenario whereby gas-phase metallicity is lowered with increased gas accretion, the similarity in gas fraction between the two samples is fully consistent with the similarity in their gas-phase metallicity. However, a small difference in gas fraction of galaxies as a function of distance to filament is observed if we restrict our analysis to $\log\left(d_{\rm skel}/\rm{Mpc}\right)<0.2$  corresponding to the decline in gas-phase metallicity close to filaments in Fig.~\ref{fig:TNG_residual}. 

The gas mass fraction as a function of stellar mass for galaxies in different distances to nodes is shown in panel (b) of Fig.~\ref{fig:TNG_gas_frac}. Galaxies closer to nodes have a lower gas fraction than those more distant from nodes with the difference becoming more pronounced at low stellar mass. In the scenario above, whereby gas accretion decreases metallicity, once again the results in Figs.~\ref{fig:TNG} and \ref{fig:TNG_gas_frac} are consistent: in that scenario, galaxies close to nodes have less access to an external gas supply, which is typically metal-poor \cite{ZReview}, and therefore experience less metal dilution, explaining their higher gas-phase metallicity.  

\subsection*{Star-formation histories and the role of halo assembly}

Simulations show that the ages of dark matter halos are correlated with the age of the stellar populations of the galaxies they host, especially at low halo or stellar mass\cite{HaloAssembly}, and star-formation histories play a role in regulating the metallicity of the gas through chemical enrichment. To investigate the role of the cosmic web on the gas-phase metallicity, we therefore look at the star-formation histories of the galaxies, in two bins of stellar mass (separated at $\log(\mstar/\msun)=$10) and in two bins of distance to the cosmic web (the 10th and 90th percentiles). We also analyse the \textit{stellar} metallicity histories, to help interpret the role of star-formation in the regulation of chemical enrichment of the galaxy. The results are shown in Fig.~\ref{fig:TNG_SFH}, with the redshift at which half the stellar mass has been formed ($z_{1/2}$) indicated for each sample. Galaxies closer to nodes, on average, formed their stars earlier with more active star-formation in the past, in direct agreement with studies that investigate halo assembly with cosmic web environment\cite{sim_bias2}. To characterise the significance of the result we perform a K-S test on the distributions of $z_{1/2}$ between the two distance bins with a significance of $7.88 \sigma$ and $3.42 \sigma$ for $\log(M_{\star}/M_{\odot})<10$ and $\log(M_{\star}/M_{\odot})>10$ respectively The impact of the earlier activity in star-formation in galaxies closer to nodes on the stellar metallicity is evident in panels (c) and (d), where galaxies closer to nodes have increased chemical enrichment. Therefore, the scenario whereby the cosmic web modulates gas-phase metallicity by modulating star-formation histories is fully consistent with our observations, provided that galaxies hold on to the enriched gas that is expelled from stars through supernovae and stellar winds.

Although not shown here, filaments have a very small impact on the star-formation histories and chemical enrichment, fitting well with the observed negligible effect of filaments on the gas-phase metallicity in Fig.~\ref{fig:TNG}(a).


\section*{Conclusions}
This study shows that the cosmic web modulates the gas-phase metallicity of galaxies. Our observational analysis has shown for the first time that the scatter in the galaxy stellar mass-gas metallicity relation (indicated by the offset in gas-phase metallicity, at fixed stellar mass, from the median relation) are significantly correlated with distance to nodes in the cosmic web, but significantly less for filaments. Crucially, we have shown that this is an effect that is specific to the anisotropic nature of the cosmic web, and not driven by overdensity. We conclude that that nodes and filaments have different roles in the scatter of the mass-metallicity relation - information that is lost if one quantifies environment by overdensity alone.

Our analysis of IllustrisTNG demonstrates that the relationship of gas-phase metallicity with distance to nodes can be explained by galaxies closer to nodes having limited access to low-metallicity gas and, at the same time, having experienced a more active star formation history in the past, meaning that they have enriched their interstellar medium with more metals than galaxies of the same mass in different cosmic web environments.
The qualitative agreement between TNG and SDSS, despite quantitative differences, is encouraging and suggests that comparing TNG and SDSS was a worthwhile exercise. However, current limitations of cosmological hydrodynamical simulations, particularly on baryonic physics, mean quantitative agreement with observations are not yet always possible. Our current work, therefore, offers a valuable opportunity in terms of model constraints: it will be informative to contrast other current and future models with our observations.

Disentangling the contributions from different physical mechanisms, and unfolding the different roles of filaments and nodes will be possible soon, by combining forthcoming statistically powerful surveys such as DESI with detailed analysis of various simulations and forward-modelled mock observations. Our work here is an important first step towards that goal. 

\newpage
\section*{Methods}

\subsection*{SDSS}
The Sloan Digital Sky Survey (SDSS)\cite{SDSS_overview} is a large galaxy imaging\cite{Gunn2008} and spectroscopic\cite{Smee2013} survey currently covering over 14,000 square degrees of sky. In this study, we use data from the DR7\cite{sdssdr7} Main Galaxy Sample\cite{Strauss2002}, which consists of a magnitude-limited sample to a petrosian r-band magnitude of $r_p = 17.77$. This sample is well suited to our work, as the survey provides a complete and dense sample of galaxies over a large contiguous area for the determination of cosmic web features, as well as high quality spectra enabling the determination of detailed galaxy properties (e.g. stellar mass, gas-phase metallicity).

\subsection*{IllustrisTNG}
IllustrisTNG is a cosmological magneto-hydrodynamical suite of simulations that models galaxy formation and evolution on cosmological scales \citep{TNG, TNG300}, by solving the coupled evolution of dark matter and baryons. The IllustrisTNG collaboration provides simulation outputs from runs performed in different volumes and at different resolutions. We mainly utilised the TNG300-1 run, corresponding to a volume of 302.6$^3$ cMpc$^3$ with a gas particle resolution of $m_{\rm baryon} = 7.6\times 10^{6} \msun/h$. This is the largest of the TNG simulations, allowing a wider variety of environments and better statistics. The smaller, higher resolution boxes were used only to check the robustness of the gas-phase metallicity relation in TNG300.

\subsection*{Catalogues and sample selections}
We use the MPA/JHU catalogue from SDSS DR7 \cite{sdssdr7} which contains stellar masses \cite{sdssms} and gas-phase metallicities \cite{SDSSMetallicities}. The stellar masses used are total stellar masses from photometry fits. The gas-phase metallicites in the catalogue were determined using a photoionisation model with the measurements of multiple emission line fluxes. The metallicities have a median 1-$\sigma$ error of 0.03 dex. In IllustrisTNG, we used the group catalogues provided by the simulation, which contain stellar masses, gas-phase metallicities and gas masses for each galaxy in snapshot 91. To better match the metallicities in SDSS, we restrict our gas-phase metallicity measurements in TNG to gas which is actively star forming. Galaxy colours for Illustris TNG were provided by a supplementary catalogue \cite{TNGColour}, from which we could extract colour-corrected colours and magnitudes.

To facilitate the comparison between observations and simulations, we performed colour and magnitude cuts in TNG. The goal here is to approximate the two samples, acknowledging that sample matching exactly is not possible nor desirable, given that simulations do not reproduce observed galaxies precisely. We broadly address sample differences due to light-cone vs snapshot, and the need to select galaxies in TNG that would have a measurable gas-phase metallicity - in SDSS, these are galaxies with emission lines strong enough to be measured in individual spectra with the required signal-to-noise. Effectively, this means removing lower-mass and redder galaxies in TNG, as well as those without gas. We selected galaxies with $-21.4<r<-18.4$ and $0.3<g-r<0.7$. The minimum stellar mass taken is $\log(\mstar/\msun)=9.25$, although the magnitude and colour cuts above effectively remove most galaxies with $\log(\mstar/\msun)<9.5$. Finally, a maximum stellar mass of $\log(\mstar/\msun)=10.2$ was chosen as we find discrepancies in the mass-metallicity relation between TNG300, TNG100 and TNG50, that seem associated with TNG300 failing to resolve relevant physical processes at high stellar mass. This is a similar range used in another study of the mass-metallicity relation in TNG \cite{mzrtng}, which showed that the MZR in IllustrisTNG is a good match to observations below that stellar mass cut-off. Our selection yielded a total sample of 97,679 galaxies in TNG300 out of a total of 225,105 galaxies above the minimum stellar mass. No cuts on SDSS DR7 were made, other than the natural selection effect of requiring gas-phase metallicity measurements. This selected 65,984 galaxies from the 302,312 in the full sample.  Further information and plots on the resulting samples are given in Supplement: Data selection with Supplementary Fig. 1 showing the colour selection on a colour-magnitude diagram and Supplementary Fig. 2 showing the resultant stellar mass distributions.


This 302,312 sample was initially selected from the full catalogue of 927,552 galaxies where the SDSS sample was restricted to galaxies within a right ascension of 120\textdegree and 220\textdegree as this provided a contiguous area allowing DisPerSE (see following section) measurements of distances to the cosmic web. The cuts described here generated a sample of SDSS 65,984 galaxies over approximately 6700 deg$^2$. 

\subsection*{Classification the cosmic web and distances to nodes and filaments}
The cosmic web was extracted using Discrete Persistent Structures Extractor (DisPerSE \cite{disperse1}, \cite{disperse2}) which measures persistent, filamentary structures in 3D. This traces out the large-scale structure using the positions of galaxies, categorising the structure into nodes, filaments, walls and voids. This allowed us to generate a catalogue of the distance to the nearest filament, $d_{\rm skel}$, and the distance to the nearest node, $d_{\rm node}$, for every galaxy in SDSS DR7 using the procedure from a recent paper\cite{disperse3}. DisPerSE was also used to generate a catalogue of the same type for every galaxy in snapshot 91 of TNG300. DisPersE measures the significance of the attachment between two points in the density field in terms of a number of standard deviations, $\sigma$. This is called the persistence level. A persistence level of 3$\sigma$ was used to generate the catalogue for SDSS DR7 and 4$\sigma$ was used for and TNG300.

In observations there is a redshift dependence on the sampling of galaxies. If distances to the cosmic web were computed without accounting for this issue, redshifts with a higher number density $n(z)$ would have artificially lower distances to nodes and filaments. This issue is typically simply resolved by normalising every distance by the mean inter-galaxy separation, $\langle D_z \rangle$, as a function of $z$: $ \langle D_z \rangle = n(z)^{-\frac{1}{3}}$. In SDSS, $ \langle D_z \rangle$ varies between approximately 4 Mpc and 12 Mpc.

When selecting the distance bins by distance to filament in SDSS, we excluded any galaxies at $d_{\rm node}<$ 1 Mpc to remove any influence from nodes and therefore isolate the relationship between metallicity and distance to filaments. This distance has been shown to be a robust cutoff for removing the influence of nodes in SDSS\cite{SDSSresults}. In TNG300 we exclude galaxies at $d_{\rm node}<$ 2 Mpc which was determined using the method described in a previous paper\cite{GalEvCW}.


\noindent
\subsection*{Metallicity residuals as a function of distance to filament/node.}


The mean mass-metallicity relations (Figs.~\ref{fig:SDSS} and~\ref{fig:TNG}) were determined using stellar mass bins of width 0.1 dex by taking the mean gas-phase metallicity in each stellar mass bin. The uncertainty of the metallicity in each bin is given as the standard error on the mean, $\sigma_{\rm mean} = \frac{\sigma_{\rm bin}}{\sqrt{N}}$, where $\sigma_{\rm bin}$ is the standard deviation of each bin and $N$ is the number of galaxies in each bin. This procedure was performed for the SDSS and IllustrisTNG samples.

The residuals of the mass-metallicity relation (Figs.~\ref{fig:SDSS_final} and~\ref{fig:TNG_residual}) were determined by taking the difference between the gas-phase metallicity of each galaxy and the mean metallicity for that stellar mass. The metallicity residuals were then binned and plotted as a function of distance to filament and node in bins of width 0.2 dex between a $\log(d_{k}/ \langle D_z \rangle)$ of -1 and 1 for SDSS and between a $\log(d_{k})$ of -1 and 1.2 in TNG (where $k$ refers to filaments or nodes). The uncertainty is given as the error on the mean, $\frac{\sigma_{\rm bin}}{\sqrt{N}}$ where $\sigma_{\rm bin}$ is the standard deviation of each bin and $N$ is the number of galaxies in each bin. 

Our interpretation of the results shown in Fig.~\ref{fig:SDSS_final} rely on quantifying how significant the correlation between gas metallicity residual and distance to the cosmic web is. We use the Spearman's rank correlation coefficient, $r_s$, to quantify the level of correlation between two variables in a non-parametric way, without assuming a linear relationship between the two (only that the relationship is monotonic). Associated with each $r_s$, we compute a p-value, that indicates the probability that a given value of $r_s$ would occur from two uncorrelated variables.
$r_s$ and p-values are computed using unbinned data.

We account for the effects of overdensity by shuffling the data with respect to overdensity. In this procedure, we construct a new binned sample such that the distribution of overdensity of the each original bin of $\log(d_{k})$ is reproduced, but we draw galaxies from the full galaxy population. The result is that each bin of our shuffled sample has an overdensity distribution that matches, by construction, the original distribution, but $\log(d_{k})$ values are no longer constrained.  We then recompute the metallicity residuals in each bin, and plot them as a function of the $\log(d_{k})$ of the bin that they were constructed to match. If the original correlations remain, we conclude the original signal was driven by overdensity, rather than distance to cosmic web features. If, on the other hand, the original correlations are significantly reduced, we conclude that overdensity did not play a major role in driving the original correlations.

Measurements of overdensity were computed using a Gaussian kernel of 3 Mpc for each galaxy in SDSS and TNG300. We also computed overdensities using a kernel size of 6 Mpc and 9 Mpc to check for potential inconsistencies in our results when using different kernel sizes. As noted in the main text, our results hold, independent of the size of Gaussian kernel. The ``shuffling" procedure was implemented in both SDSS and TNG metallicity residuals. In addition, in TNG we leveraged the larger sample size and more robust measurements of overdensity to also demonstrate the effect of distances to cosmic web in a very small bin of overdensity, as shown in Fig.~\ref{fig:TNG}.

\subsection*{Star formation and chemical enrichment histories in IllustrisTNG.}

In IllustrisTNG we extract the ages and metallicity of all stellar particles bound to each subhalo in our sample, up to two effective radii. The star-formation histories are computed as the mass formed in each of 400 bins of lookback time, linearly spaced between 0 and the age of the Universe at $z=0.1$. The metallicity histories are computed by taking the the mass-weighted metallicity in the same bins.


\bibliographystyle{naturemag}
\bibliography{sample} 

\begin{thebibliography}{10}
\expandafter\ifx\csname url\endcsname\relax
  \def\url#1{\texttt{#1}}\fi
\expandafter\ifx\csname urlprefix\endcsname\relax\def\urlprefix{URL }\fi
\providecommand{\bibinfo}[2]{#2}
\providecommand{\eprint}[2][]{\url{#2}}

\bibitem{WechslerTinker2018}
\bibinfo{author}{{Wechsler}, R.~H.} \& \bibinfo{author}{{Tinker}, J.~L.}
\newblock \bibinfo{title}{{The Connection Between Galaxies and Their Dark
  Matter Halos}}.
\newblock \emph{\bibinfo{journal}{\araa}} \textbf{\bibinfo{volume}{56}},
  \bibinfo{pages}{435--487} (\bibinfo{year}{2018}).
\newblock \eprint{1804.03097}.

\bibitem{BlantonMoustakas2009}
\bibinfo{author}{{Blanton}, M.~R.} \& \bibinfo{author}{{Moustakas}, J.}
\newblock \bibinfo{title}{{Physical Properties and Environments of Nearby
  Galaxies}}.
\newblock \emph{\bibinfo{journal}{\araa}} \textbf{\bibinfo{volume}{47}},
  \bibinfo{pages}{159--210} (\bibinfo{year}{2009}).
\newblock \eprint{0908.3017}.

\bibitem{cosmicweb}
\bibinfo{author}{{Bond}, J.~R.}, \bibinfo{author}{{Kofman}, L.} \&
  \bibinfo{author}{{Pogosyan}, D.}
\newblock \bibinfo{title}{{How filaments of galaxies are woven into the cosmic
  web}}.
\newblock \emph{\bibinfo{journal}{\nat}} \textbf{\bibinfo{volume}{380}},
  \bibinfo{pages}{603--606} (\bibinfo{year}{1996}).
\newblock \eprint{astro-ph/9512141}.

\bibitem{sim_bias1}
\bibinfo{author}{{Dalal}, N.}, \bibinfo{author}{{White}, M.},
  \bibinfo{author}{{Bond}, J.~R.} \& \bibinfo{author}{{Shirokov}, A.}
\newblock \bibinfo{title}{{Halo Assembly Bias in Hierarchical Structure
  Formation}}.
\newblock \emph{\bibinfo{journal}{\apj}} \textbf{\bibinfo{volume}{687}},
  \bibinfo{pages}{12--21} (\bibinfo{year}{2008}).
\newblock \eprint{0803.3453}.

\bibitem{sim_bias2}
\bibinfo{author}{{Borzyszkowski}, M.}, \bibinfo{author}{{Porciani}, C.},
  \bibinfo{author}{{Romano-D{\'\i}az}, E.} \& \bibinfo{author}{{Garaldi}, E.}
\newblock \bibinfo{title}{{ZOMG - I. How the cosmic web inhibits halo growth
  and generates assembly bias}}.
\newblock \emph{\bibinfo{journal}{\mnras}} \textbf{\bibinfo{volume}{469}},
  \bibinfo{pages}{594--611} (\bibinfo{year}{2017}).
\newblock \eprint{1610.04231}.

\bibitem{sim_bias3}
\bibinfo{author}{{Musso}, M.} \emph{et~al.}
\newblock \bibinfo{title}{{How does the cosmic web impact assembly bias?}}
\newblock \emph{\bibinfo{journal}{\mnras}} \textbf{\bibinfo{volume}{476}},
  \bibinfo{pages}{4877--4906} (\bibinfo{year}{2018}).
\newblock \eprint{1709.00834}.

\bibitem{sim_bias4}
\bibinfo{author}{{Paranjape}, A.}, \bibinfo{author}{{Hahn}, O.} \&
  \bibinfo{author}{{Sheth}, R.~K.}
\newblock \bibinfo{title}{{Halo assembly bias and the tidal anisotropy of the
  local halo environment}}.
\newblock \emph{\bibinfo{journal}{\mnras}} \textbf{\bibinfo{volume}{476}},
  \bibinfo{pages}{3631--3647} (\bibinfo{year}{2018}).
\newblock \eprint{1706.09906}.

\bibitem{HaloAssembly}
\bibinfo{author}{{Tojeiro}, R.} \emph{et~al.}
\newblock \bibinfo{title}{{Galaxy and Mass Assembly (GAMA): halo formation
  times and halo assembly bias on the cosmic web}}.
\newblock \emph{\bibinfo{journal}{\mnras}} \textbf{\bibinfo{volume}{470}},
  \bibinfo{pages}{3720--3741} (\bibinfo{year}{2017}).
\newblock \eprint{1612.08595}.

\bibitem{GalEvCW3}
\bibinfo{author}{{Alpaslan}, M.} \emph{et~al.}
\newblock \bibinfo{title}{{Galaxy And Mass Assembly (GAMA): stellar mass growth
  of spiral galaxies in the cosmic web}}.
\newblock \emph{\bibinfo{journal}{\mnras}} \textbf{\bibinfo{volume}{457}},
  \bibinfo{pages}{2287--2300} (\bibinfo{year}{2016}).
\newblock \eprint{1601.03391}.

\bibitem{GalEvCW}
\bibinfo{author}{{Kraljic}, K.} \emph{et~al.}
\newblock \bibinfo{title}{{Galaxy evolution in the metric of the cosmic web}}.
\newblock \emph{\bibinfo{journal}{\mnras}} \textbf{\bibinfo{volume}{474}},
  \bibinfo{pages}{547--571} (\bibinfo{year}{2018}).
\newblock \eprint{1710.02676}.

\bibitem{SDSSresults}
\bibinfo{author}{{Winkel}, N.} \emph{et~al.}
\newblock \bibinfo{title}{{The imprint of cosmic web quenching on central
  galaxies}}.
\newblock \emph{\bibinfo{journal}{\mnras}} \textbf{\bibinfo{volume}{505}},
  \bibinfo{pages}{4920--4934} (\bibinfo{year}{2021}).
\newblock \eprint{2105.13368}.

\bibitem{HI_replenish}
\bibinfo{author}{{Kleiner}, D.}, \bibinfo{author}{{Pimbblet}, K.~A.},
  \bibinfo{author}{{Jones}, D.~H.}, \bibinfo{author}{{Koribalski}, B.~S.} \&
  \bibinfo{author}{{Serra}, P.}
\newblock \bibinfo{title}{{Evidence for H I replenishment in massive galaxies
  through gas accretion from the cosmic web}}.
\newblock \emph{\bibinfo{journal}{\mnras}} \textbf{\bibinfo{volume}{466}},
  \bibinfo{pages}{4692--4710} (\bibinfo{year}{2017}).
\newblock \eprint{1701.03467}.

\bibitem{HI}
\bibinfo{author}{{Crone Odekon}, M.} \emph{et~al.}
\newblock \bibinfo{title}{{The Effect of Filaments and Tendrils on the H I
  Content of Galaxies}}.
\newblock \emph{\bibinfo{journal}{\apj}} \textbf{\bibinfo{volume}{852}},
  \bibinfo{pages}{142} (\bibinfo{year}{2018}).
\newblock \eprint{1712.05045}.

\bibitem{Darvish+15}
\bibinfo{author}{{Darvish}, B.} \emph{et~al.}
\newblock \bibinfo{title}{{Spectroscopic Study of Star-forming Galaxies in
  Filaments and the Field at z \raisebox{-0.5ex}\textasciitilde 0.5: Evidence
  for Environmental Dependence of Electron Density}}.
\newblock \emph{\bibinfo{journal}{\apj}} \textbf{\bibinfo{volume}{814}},
  \bibinfo{pages}{84} (\bibinfo{year}{2015}).
\newblock \eprint{1510.05009}.

\bibitem{Genel16}
\bibinfo{author}{{Genel}, S.}
\newblock \bibinfo{title}{{How Environment Affects Galaxy Metallicity through
  Stripping and Formation History: Lessons from the Illustris Simulation}}.
\newblock \emph{\bibinfo{journal}{\apj}} \textbf{\bibinfo{volume}{822}},
  \bibinfo{pages}{107} (\bibinfo{year}{2016}).
\newblock \eprint{1602.02773}.

\bibitem{Gupta+18}
\bibinfo{author}{{Gupta}, A.} \emph{et~al.}
\newblock \bibinfo{title}{{Chemical pre-processing of cluster galaxies over the
  past 10 billion years in the IllustrisTNG simulations}}.
\newblock \emph{\bibinfo{journal}{\mnras}} \textbf{\bibinfo{volume}{477}},
  \bibinfo{pages}{L35--L39} (\bibinfo{year}{2018}).
\newblock \eprint{1801.03500}.

\bibitem{ZReview}
\bibinfo{author}{{Maiolino}, R.} \& \bibinfo{author}{{Mannucci}, F.}
\newblock \bibinfo{title}{{De re metallica: the cosmic chemical evolution of
  galaxies}}.
\newblock \emph{\bibinfo{journal}{\aapr}} \textbf{\bibinfo{volume}{27}},
  \bibinfo{pages}{3} (\bibinfo{year}{2019}).
\newblock \eprint{1811.09642}.

\bibitem{SDSSMetallicities}
\bibinfo{author}{{Tremonti}, C.~A.} \emph{et~al.}
\newblock \bibinfo{title}{{The Origin of the Mass-Metallicity Relation:
  Insights from 53,000 Star-forming Galaxies in the Sloan Digital Sky Survey}}.
\newblock \emph{\bibinfo{journal}{\apj}} \textbf{\bibinfo{volume}{613}},
  \bibinfo{pages}{898--913} (\bibinfo{year}{2004}).
\newblock \eprint{astro-ph/0405537}.

\bibitem{scatterMZR_2}
\bibinfo{author}{{De Lucia}, G.}, \bibinfo{author}{{Xie}, L.},
  \bibinfo{author}{{Fontanot}, F.} \& \bibinfo{author}{{Hirschmann}, M.}
\newblock \bibinfo{title}{{Gas accretion regulates the scatter of the
  mass-metallicity relation}}.
\newblock \emph{\bibinfo{journal}{\mnras}} \textbf{\bibinfo{volume}{498}},
  \bibinfo{pages}{3215--3227} (\bibinfo{year}{2020}).
\newblock \eprint{2008.09127}.

\bibitem{scatterMZR}
\bibinfo{author}{{van Loon}, M.~L.}, \bibinfo{author}{{Mitchell}, P.~D.} \&
  \bibinfo{author}{{Schaye}, J.}
\newblock \bibinfo{title}{{Explaining the scatter in the galaxy
  mass-metallicity relation with gas flows}}.
\newblock \emph{\bibinfo{journal}{\mnras}} \textbf{\bibinfo{volume}{504}},
  \bibinfo{pages}{4817--4828} (\bibinfo{year}{2021}).
\newblock \eprint{2101.11021}.

\bibitem{MannucciEtAl2010}
\bibinfo{author}{{Mannucci}, F.}, \bibinfo{author}{{Cresci}, G.},
  \bibinfo{author}{{Maiolino}, R.}, \bibinfo{author}{{Marconi}, A.} \&
  \bibinfo{author}{{Gnerucci}, A.}
\newblock \bibinfo{title}{{A fundamental relation between mass, star formation
  rate and metallicity in local and high-redshift galaxies}}.
\newblock \emph{\bibinfo{journal}{\mnras}} \textbf{\bibinfo{volume}{408}},
  \bibinfo{pages}{2115--2127} (\bibinfo{year}{2010}).
\newblock \eprint{1005.0006}.

\bibitem{DESI}
\bibinfo{author}{{Martini}, P.} \emph{et~al.}
\newblock \bibinfo{title}{{Overview of the Dark Energy Spectroscopic
  Instrument}}.
\newblock In \bibinfo{editor}{{Evans}, C.~J.}, \bibinfo{editor}{{Simard}, L.}
  \& \bibinfo{editor}{{Takami}, H.} (eds.)
  \emph{\bibinfo{booktitle}{Ground-based and Airborne Instrumentation for
  Astronomy VII}}, vol. \bibinfo{volume}{10702} of
  \emph{\bibinfo{series}{Society of Photo-Optical Instrumentation Engineers
  (SPIE) Conference Series}}, \bibinfo{pages}{107021F} (\bibinfo{year}{2018}).
\newblock \eprint{1807.09287}.

\bibitem{SDSS_overview}
\bibinfo{author}{{York}, D.~G.} \emph{et~al.}
\newblock \bibinfo{title}{{The Sloan Digital Sky Survey: Technical Summary}}.
\newblock \emph{\bibinfo{journal}{\aj}} \textbf{\bibinfo{volume}{120}},
  \bibinfo{pages}{1579--1587} (\bibinfo{year}{2000}).
\newblock \eprint{astro-ph/0006396}.

\bibitem{sdssdr7}
\bibinfo{author}{{Abazajian}, K.~N.} \emph{et~al.}
\newblock \bibinfo{title}{{The Seventh Data Release of the Sloan Digital Sky
  Survey}}.
\newblock \emph{\bibinfo{journal}{\apjs}} \textbf{\bibinfo{volume}{182}},
  \bibinfo{pages}{543--558} (\bibinfo{year}{2009}).
\newblock \eprint{0812.0649}.

\bibitem{Strauss2002}
\bibinfo{author}{{Strauss}, M.~A.} \emph{et~al.}
\newblock \bibinfo{title}{{Spectroscopic Target Selection in the Sloan Digital
  Sky Survey: The Main Galaxy Sample}}.
\newblock \emph{\bibinfo{journal}{\aj}} \textbf{\bibinfo{volume}{124}},
  \bibinfo{pages}{1810--1824} (\bibinfo{year}{2002}).
\newblock \eprint{astro-ph/0206225}.

\bibitem{localMetal}
\bibinfo{author}{{Peng}, Y.-j.} \& \bibinfo{author}{{Maiolino}, R.}
\newblock \bibinfo{title}{{The dependence of the galaxy mass-metallicity
  relation on environment and the implied metallicity of the IGM}}.
\newblock \emph{\bibinfo{journal}{\mnras}} \textbf{\bibinfo{volume}{438}},
  \bibinfo{pages}{262--270} (\bibinfo{year}{2014}).
\newblock \eprint{1311.1816}.

\bibitem{localMetal2}
\bibinfo{author}{{Chartab}, N.} \emph{et~al.}
\newblock \bibinfo{title}{{The MOSDEF Survey: Environmental Dependence of the
  Gas-phase Metallicity of Galaxies at 1.4 {\ensuremath{\leq}} z
  {\ensuremath{\leq}} 2.6}}.
\newblock \emph{\bibinfo{journal}{\apj}} \textbf{\bibinfo{volume}{908}},
  \bibinfo{pages}{120} (\bibinfo{year}{2021}).
\newblock \eprint{2101.01706}.

\bibitem{localMetal3}
\bibinfo{author}{{Williams}, R.~J.} \emph{et~al.}
\newblock \bibinfo{title}{{Dynamics and metallicity of far-infrared selected
  galaxies}}.
\newblock \emph{\bibinfo{journal}{\mnras}} \textbf{\bibinfo{volume}{443}},
  \bibinfo{pages}{3780--3794} (\bibinfo{year}{2014}).
\newblock \eprint{1405.3664}.

\bibitem{Libeskind2018}
\bibinfo{author}{{Libeskind}, N.~I.} \emph{et~al.}
\newblock \bibinfo{title}{{Tracing the cosmic web}}.
\newblock \emph{\bibinfo{journal}{\mnras}} \textbf{\bibinfo{volume}{473}},
  \bibinfo{pages}{1195--1217} (\bibinfo{year}{2018}).
\newblock \eprint{1705.03021}.

\bibitem{TNGColour}
\bibinfo{author}{{Nelson}, D.} \emph{et~al.}
\newblock \bibinfo{title}{{First results from the IllustrisTNG simulations: the
  galaxy colour bimodality}}.
\newblock \emph{\bibinfo{journal}{\mnras}} \textbf{\bibinfo{volume}{475}},
  \bibinfo{pages}{624--647} (\bibinfo{year}{2018}).
\newblock \eprint{1707.03395}.

\bibitem{TNG}
\bibinfo{author}{{Pillepich}, A.} \emph{et~al.}
\newblock \bibinfo{title}{{Simulating galaxy formation with the IllustrisTNG
  model}}.
\newblock \emph{\bibinfo{journal}{\mnras}} \textbf{\bibinfo{volume}{473}},
  \bibinfo{pages}{4077--4106} (\bibinfo{year}{2018}).
\newblock \eprint{1703.02970}.

\bibitem{TNG300}
\bibinfo{author}{{Springel}, V.} \emph{et~al.}
\newblock \bibinfo{title}{{First results from the IllustrisTNG simulations:
  matter and galaxy clustering}}.
\newblock \emph{\bibinfo{journal}{\mnras}} \textbf{\bibinfo{volume}{475}},
  \bibinfo{pages}{676--698} (\bibinfo{year}{2018}).
\newblock \eprint{1707.03397}.

\bibitem{CGM}
\bibinfo{author}{{Tumlinson}, J.}, \bibinfo{author}{{Peeples}, M.~S.} \&
  \bibinfo{author}{{Werk}, J.~K.}
\newblock \bibinfo{title}{{The Circumgalactic Medium}}.
\newblock \emph{\bibinfo{journal}{\araa}} \textbf{\bibinfo{volume}{55}},
  \bibinfo{pages}{389--432} (\bibinfo{year}{2017}).
\newblock \eprint{1709.09180}.

\bibitem{Gunn2008}
\bibinfo{author}{{Gunn}, J.~E.} \emph{et~al.}
\newblock \bibinfo{title}{{The Sloan Digital Sky Survey Photometric Camera}}.
\newblock \emph{\bibinfo{journal}{\aj}} \textbf{\bibinfo{volume}{116}},
  \bibinfo{pages}{3040--3081} (\bibinfo{year}{1998}).
\newblock \eprint{astro-ph/9809085}.

\bibitem{Smee2013}
\bibinfo{author}{{Smee}, S.~A.} \emph{et~al.}
\newblock \bibinfo{title}{{The Multi-object, Fiber-fed Spectrographs for the
  Sloan Digital Sky Survey and the Baryon Oscillation Spectroscopic Survey}}.
\newblock \emph{\bibinfo{journal}{\aj}} \textbf{\bibinfo{volume}{146}},
  \bibinfo{pages}{32} (\bibinfo{year}{2013}).
\newblock \eprint{1208.2233}.

\bibitem{sdssms}
\bibinfo{author}{{Salim}, S.} \emph{et~al.}
\newblock \bibinfo{title}{{UV Star Formation Rates in the Local Universe}}.
\newblock \emph{\bibinfo{journal}{\apjs}} \textbf{\bibinfo{volume}{173}},
  \bibinfo{pages}{267--292} (\bibinfo{year}{2007}).
\newblock \eprint{0704.3611}.

\bibitem{mzrtng}
\bibinfo{author}{{Torrey}, P.} \emph{et~al.}
\newblock \bibinfo{title}{{The evolution of the mass-metallicity relation and
  its scatter in IllustrisTNG}}.
\newblock \emph{\bibinfo{journal}{\mnras}} \textbf{\bibinfo{volume}{484}},
  \bibinfo{pages}{5587--5607} (\bibinfo{year}{2019}).
\newblock \eprint{1711.05261}.

\bibitem{disperse1}
\bibinfo{author}{{Sousbie}, T.}
\newblock \bibinfo{title}{{The persistent cosmic web and its filamentary
  structure - I. Theory and implementation}}.
\newblock \emph{\bibinfo{journal}{\mnras}} \textbf{\bibinfo{volume}{414}},
  \bibinfo{pages}{350--383} (\bibinfo{year}{2011}).
\newblock \eprint{1009.4015}.

\bibitem{disperse2}
\bibinfo{author}{{Sousbie}, T.}, \bibinfo{author}{{Pichon}, C.} \&
  \bibinfo{author}{{Kawahara}, H.}
\newblock \bibinfo{title}{{The persistent cosmic web and its filamentary
  structure - II. Illustrations}}.
\newblock \emph{\bibinfo{journal}{\mnras}} \textbf{\bibinfo{volume}{414}},
  \bibinfo{pages}{384--403} (\bibinfo{year}{2011}).
\newblock \eprint{1009.4014}.

\bibitem{disperse3}
\bibinfo{author}{{Kraljic}, K.} \emph{et~al.}
\newblock \bibinfo{title}{{The impact of the connectivity of the cosmic web on
  the physical properties of galaxies at its nodes}}.
\newblock \emph{\bibinfo{journal}{\mnras}} \textbf{\bibinfo{volume}{491}},
  \bibinfo{pages}{4294--4309} (\bibinfo{year}{2020}).
\newblock \eprint{1910.08066}.

\end{thebibliography}

\noindent
\textbf{Data Availability}
Data for SDSS DR7 and Illustris TNG are publicly available at the respective links: \\ https://wwwmpa.mpa-garching.mpg.de/SDSS/DR7/,\\ https://www.tng-project.org/data/. \\ DisPerSE catalogues are available upon reasonable request. \\
\noindent
\textbf{Code Availability}
DisPerSE is publicly available at: 

http://www2.iap.fr/users/sousbie/web/html/indexba87.html?category/Install. 

All other code used in this project is available upon reasonable request. \\
\noindent
\textbf{Acknowledgements}
Part of this work has made use of the Horizon cluster hosted by the Institut d'Astrophysique de Paris. We warmly thank S.~Rouberol for running it smoothly. KK acknowledges support from the DEEPDIP project (ANR-19-CE31-0023).

Funding for the Sloan Digital Sky Survey IV has been provided by the Alfred P. Sloan Foundation, the U.S. Department of Energy Office of Science, and the Participating Institutions. SDSS acknowledges support and resources from the Center for High-Performance Computing at the University of Utah. The SDSS web site is www.sdss.org.

SDSS is managed by the Astrophysical Research Consortium for the Participating Institutions of the SDSS Collaboration including the Brazilian Participation Group, the Carnegie Institution for Science, Carnegie Mellon University, Center for Astrophysics | Harvard \& Smithsonian (CfA), the Chilean Participation Group, the French Participation Group, Instituto de Astrofísica de Canarias, The Johns Hopkins University, Kavli Institute for the Physics and Mathematics of the Universe (IPMU) / University of Tokyo, the Korean Participation Group, Lawrence Berkeley National Laboratory, Leibniz Institut für Astrophysik Potsdam (AIP), Max-Planck-Institut für Astronomie (MPIA Heidelberg), Max-Planck-Institut für Astrophysik (MPA Garching), Max-Planck-Institut für Extraterrestrische Physik (MPE), National Astronomical Observatories of China, New Mexico State University, New York University, University of Notre Dame, Observatório Nacional / MCTI, The Ohio State University, Pennsylvania State University, Shanghai Astronomical Observatory, United Kingdom Participation Group, Universidad Nacional Autónoma de México, University of Arizona, University of Colorado Boulder, University of Oxford, University of Portsmouth, University of Utah, University of Virginia, University of Washington, University of Wisconsin, Vanderbilt University, and Yale University.

\noindent
\textbf{Author Contributions}
CTD performed the main analysis of the data. CTD, RT, KK interpreted the results and contributed to the writing of the manuscript. KK generated the DisPerSE catalogues.  \\
\noindent
\textbf{Corresponding Author} Correspondence to Callum T. Donnan (email: callum.donnan@ed.ac.uk).

\noindent
\textbf{Competing Interests} The authors declare no competing interests.

\newpage

\section*{Supplement: Data selection}
\begin{figure}[H]
    \centering
    \includegraphics[width=\columnwidth]{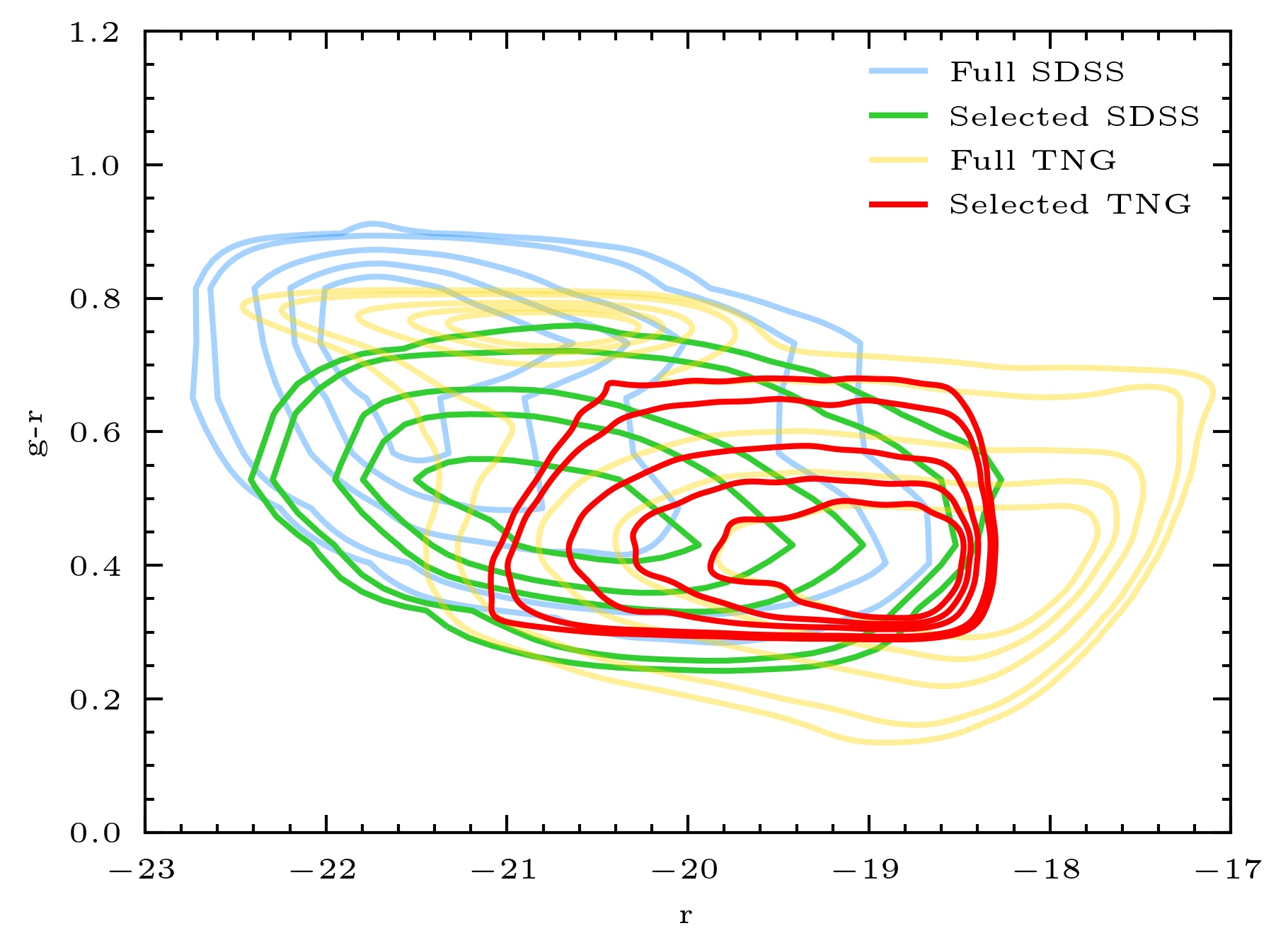}
    \caption{\textbf{Colour-magnitude diagram for SDSS DR7 and TNG300.} Contains each full dataset with the corresponding data selection used in this paper. r-band magnitudes are shown as rest-frame absolute magnitudes. Contours represent 2D kernel density estimates at \{0.05,0.1,0.3,0.5,0.7\}. }
    \label{fig:CM}
\end{figure}

\begin{figure}[H]
    \centering
    \includegraphics[width=\columnwidth]{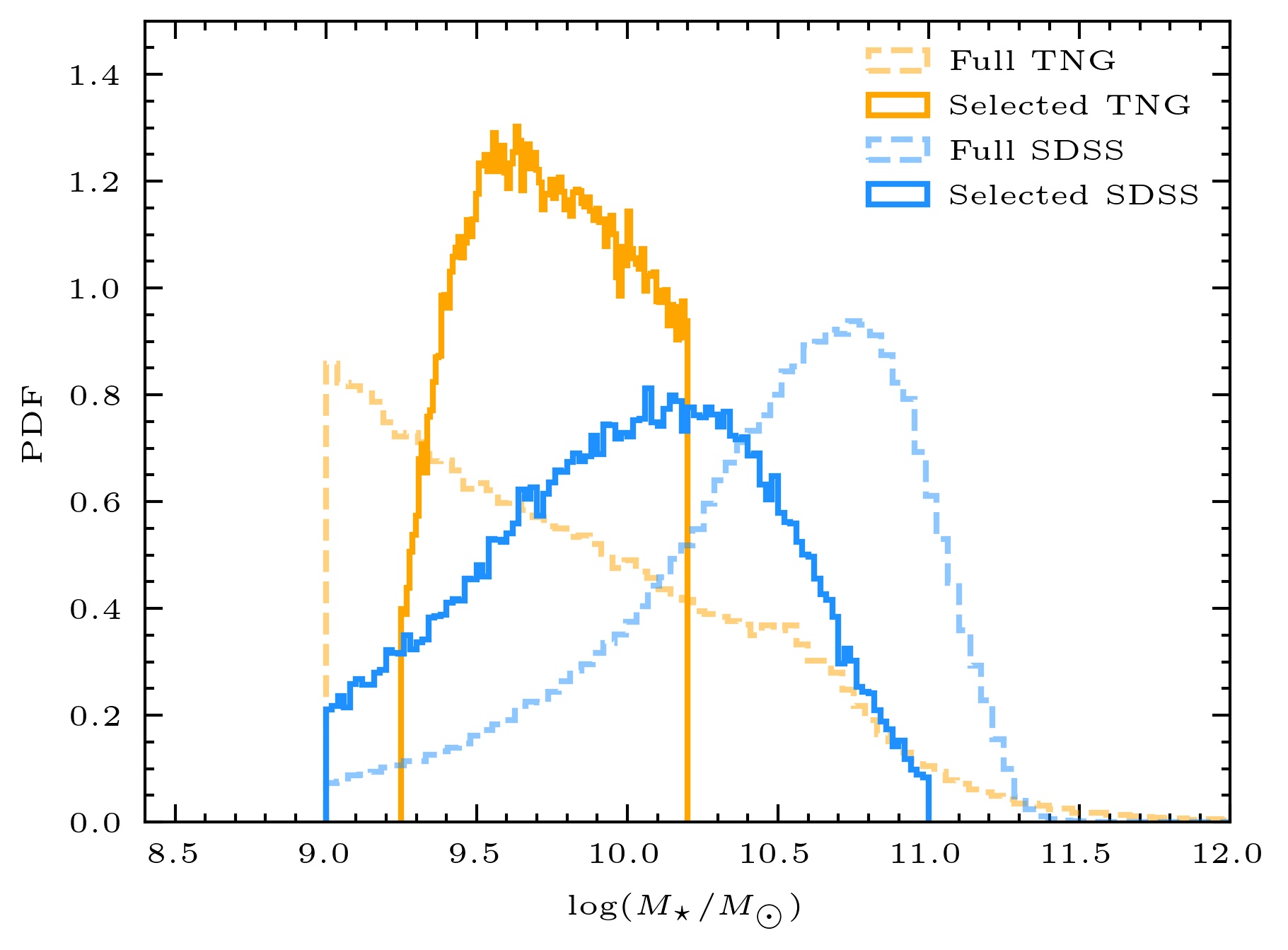}
    \caption{\textbf{Stellar mass distributions in SDSS and TNG300 with sample selection.} Stellar mass distributions are normalised and are shown for the full SDSS DR7 sample, full TNG300 (at $z$= 0.1) sample, selected SDSS DR7 sample and selected TNG300 sample in 100 bins. }
    \label{fig:stellar_mass_dist}
\end{figure}

Supplementary Fig.~\ref{fig:CM} shows the colour-magnitude plot of $g$-$r$ colour against the r-band absolute magnitude for SDSS and TNG300. The blue contours illustrate the full sample in DR7 of SDSS with the green contours representing our sample selection: galaxies in the catalogue with a gas-phase metallicity measurement. The yellow contours represent the full TNG300 sample with the red contours illustrating the selection as described in the methods section. Supplementary Fig.~\ref{fig:stellar_mass_dist} shows the resultant stellar mass distributions for the sample selection in SDSS and TNG300, illustrating how the selections made, better match the two samples than simply considering the full dataset for both.

\newpage
\section*{Supplement: Metallicity residuals as function of overdensity}
In Fig.~2 we showed that the relationship between gas-phase metallicity and the cosmic web was independent of overdensity. However, it is important to note that although this relationship was independent of overdensity, we still recover the correlation between gas-phase metallicity and overdensity that has been reported in the literature\textsuperscript{26;27;28} in our analysis. This is shown in Supplementary Fig.~\ref{fig:overdens_residual}. Galaxies at all stellar masses considered are enriched in local overdensities with this effect being more pronounced at low stellar mass.
\begin{figure}[H]
    \centering
    \includegraphics[width=\columnwidth]{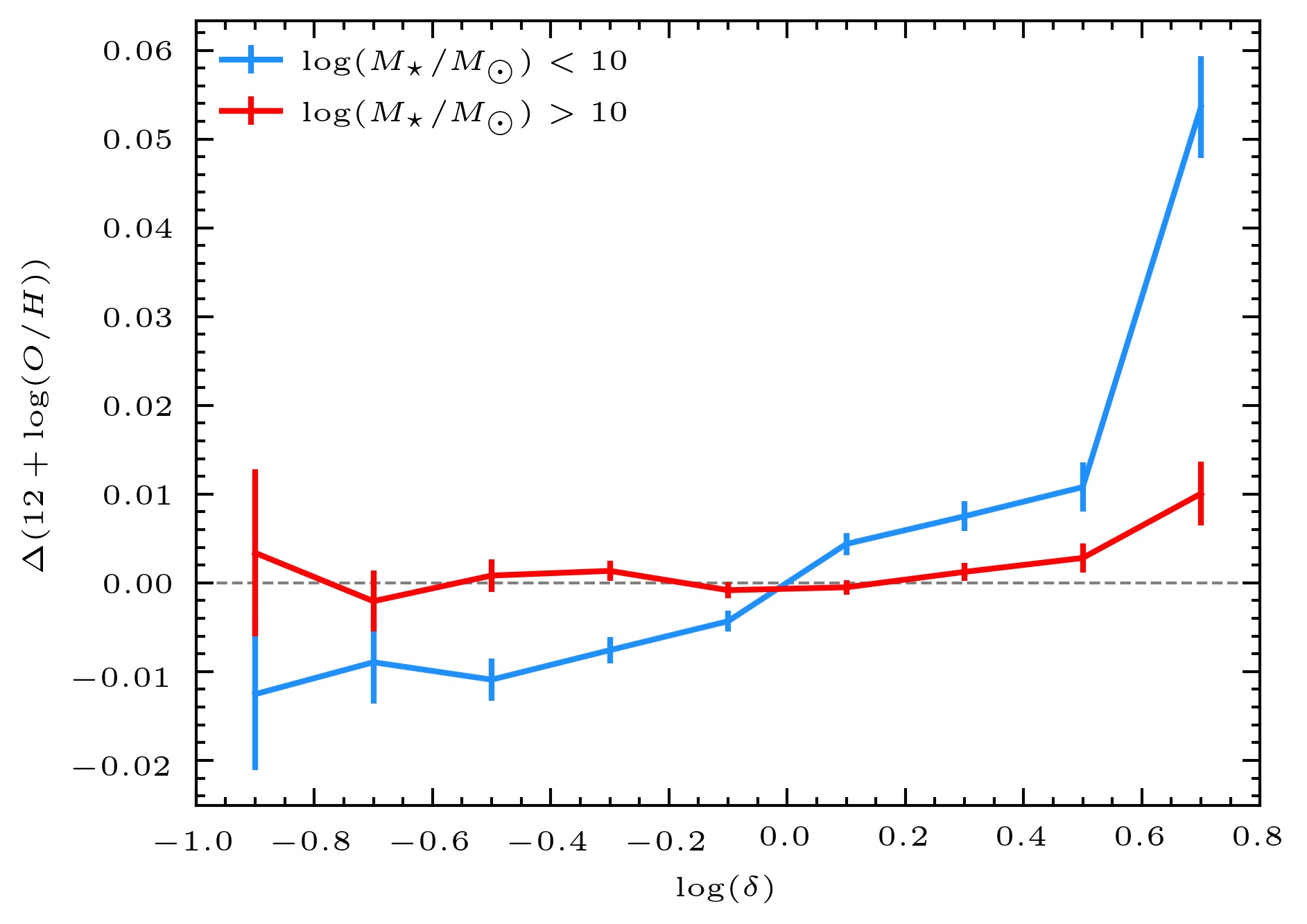}
    \caption{\textbf{Binned gas-phase metallicity residuals as a function of overdensity in SDSS DR7.} Residuals were binned after seperating the samples by a stellar mass cut at $\log(M_{\star}/M_{\odot})=10$.}
    \label{fig:overdens_residual}
\end{figure}
\end{document}